\documentstyle[prl,amsfonts,aps]{revtex}
\begin{document}
\title{\bf Addendum to \\ ``Multichannel Kondo screening in a
one-dimensional correlated electron system''}
\vspace{1.0em}
\author{A.~ A.~Zvyagin$^{(a,b)}$, H.~Johannesson$^{(a)}$ and M. Granath$^{(a)}$}
\address{$^{(a)}$Institute of Theoretical Physics, Chalmers
University of Technology and G\"oteborg University, SE 412 96
G\"oteborg,
Sweden}
\address{$^{(b)}$B.~I.~Verkin Institute for Low Temperature Physics
and Engineering \\ of the National Ukrainian Academy of Sciences,
310164
Kharkov, Ukraine}
\maketitle
\begin{abstract}
This is an addendum to our previous work
{\tt cond-mat/9705048} (published in
Europhysics Letters 41, 213 (1998)), clarifying the construction of
the two-particle scattering matrices used for studying the magnetic
impurity behavior in a multichannel correlated host. \\ \\
Published in {\it Europhysics Letters {\bf 50}, 125 (2000)}. \\ \\ \\
\end{abstract}

We point out that our lattice host Hamiltonian in Eq. (1), {\tt cond-mat/9705048},
(Europhysics Letters 41, 213 (1998)) 
does not {\it directly}
produce the electron-electron scattering matrix ${\hat X}(p_i-p_j)$ as given in the
subsequent text.
To obtain the factorizable form of this matrix
one must first {\it linearize the dispersion relation} of
the itinerant electrons about the Fermi levels (with the Fermi
velocity normalized to unity). The linearized impurity Hamiltonian
is constructed from the lattice form, Eq. (3), in a similar way (for
details, see P. Schlottmann and A. A. Zvyagin, Phys. Rev. B, {\bf 55}, 5027 (1997)).
With a linearized spectrum all our results for the behavior of the magnetic
impurity are ensured to be valid.
\end{document}